\documentclass[aip, reprint, superscriptaddress,nofootinbib]{revtex4-1}
\usepackage{color}
\usepackage{graphicx}
\usepackage{siunitx}
\usepackage{pgfplots}
\usepackage{physics}
\usepackage{hyperref}
\usepackage{ulem}

\usetikzlibrary{calc}
\begin{document}

\title{Coherent Remote Control of Quantum Emitters Embedded in Polymer Waveguides}
\author{Alexander Landowski}
\email[Both authors contributed equally]{}
\email[email:]{gutsche@rhrk.uni-kl.de}
\affiliation{Department of Physics and State Research Center OPTIMAS, University of Kaiserslautern, Erwin-Schroedinger-Str. 46, 67663 Kaiserslautern, Germany}
\author{Jonas Gutsche}
\email[Both authors contributed equally]{}
\email[email:]{gutsche@rhrk.uni-kl.de}
\affiliation{Department of Physics and State Research Center OPTIMAS, University of Kaiserslautern, Erwin-Schroedinger-Str. 46, 67663 Kaiserslautern, Germany}
\affiliation{Graduate School Materials Science in Mainz, Erwin-Schroedinger-Str. 46, 67663 Kaiserslautern, Germany}
\author{Stefan Guckenbiehl}
\affiliation{Department of Physics and State Research Center OPTIMAS, University of Kaiserslautern, Erwin-Schroedinger-Str. 46, 67663 Kaiserslautern, Germany}
\author{Marius Sch\"onberg}
\affiliation{Department of Physics and State Research Center OPTIMAS, University of Kaiserslautern, Erwin-Schroedinger-Str. 46, 67663 Kaiserslautern, Germany}
\author{Georg von Freymann}
\affiliation{Department of Physics and State Research Center OPTIMAS, University of Kaiserslautern, Erwin-Schroedinger-Str. 46, 67663 Kaiserslautern, Germany}
\affiliation{Fraunhofer Institute for Industrial Mathematics ITWM, Fraunhofer-Platz 1, 67663 Kaiserslautern, Germany}
\author{Artur Widera}
\affiliation{Department of Physics and State Research Center OPTIMAS, University of Kaiserslautern, Erwin-Schroedinger-Str. 46, 67663 Kaiserslautern, Germany}
\affiliation{Graduate School Materials Science in Mainz, Erwin-Schroedinger-Str. 46, 67663 Kaiserslautern, Germany}

\begin{abstract}
We report on the coherent internal-state control of single crystalline nanodiamonds, containing on average 1200 nitrogen-vacancy (NV) centers, embedded in three-dimensional direct-laser-written waveguides. We excite the NV centers by light propagating through the waveguide, and we show that emitted fluorescence can be efficiently coupled to the waveguide modes. 
We find an average coupling efficiency of \SI{21.6}{\percent} into all guided modes.
Moreover, we investigate optically-detected magnetic-resonance spectra as well as Rabi oscillations recorded through the waveguide-coupled signal. Our work shows that the system is well suited for magnetometry and remote read-out of spin coherence in a freely configurable waveguide network, overcoming the need for direct optical access of NV centers in nanodiamonds.
These waveguide-integrated sensors might open up new applications, like determining magnetic field distributions inside opaque or scattering media, or photosensitive samples, like biological tissue.
\end{abstract}

\maketitle

\section{Introduction}
Nitrogen-vacancy (NV) centers in diamond have advanced to a highly promising nano-scale probe. 
A prominent feature of the NV center is optical initialization and readout of its spin degree of freedom \cite{Jelezko.2004}. 
Because of relatively long coherence times, for example up to $T_2=\SI{3}{\milli\second}$ at room temperature \cite{BarGill.2013}, and its high sensitivity to external fields, it is widely used to detect DC and AC magnetic fields \cite{Balasubramanian.2008,Maze.2008b,Grinolds.2013,Sushkov.2014}, to sense temperature distributions in biological samples \cite{Kucsko.2013}, and even to show a loophole free test of the Bell inequality \cite{Hensen.2015}.

Most such applications require a direct line of sight to the location of the NV center investigated for excitation and fluorescence light detection, \textit{i.e.} direct optical access, usually granted by microscopes with high numerical apertures. An alternative way to guide the NV center's signal to a detector are waveguides. 
Such waveguides, directly fabricated from a diamond membrane \cite{Babinec.2010b,Maletinsky.2012,Momenzadeh.2015,Mouradian.2015} or micrometer long, free standing polymer waveguides \cite{Schell.2013,Shi.2016b}, guiding NV fluorescence signal have been reported elsewhere.
Moreover, it has been shown that waveguides as well as NV centers can be deterministically defined inside bulk diamond using a direct laser writing technique \cite{Hadden.2018}.
\\
Here, we show the integration of individual single-crystalline nanodiamonds containing an ensemble of NV centers into flexible, three-dimensional waveguide networks that can be on millimeter length scales and connected to more complex optical networks with high integration density. 
Our motivation to choose polymer waveguides is threefold.
First, we position the nanodiamond inside the waveguide, where the intensity of the guided modes is much higher than in the evanescent field outside the waveguide. 
This allows for remote excitation as well as detection of the NV centers' fluorescence and thus ensures constant and optimal coupling of light to and from the NV centers independent of external conditions. 
Second, polymer waveguides feature 3D capability, which enables simultaneous addressing of NV center ensembles lying in different focal planes, while keeping one side of the sample accessible for additional manipulation.
These 3D capabilities will also allow for integration into microoptics or microfluidics on a chip \cite{Ziem.2013}.
Third, such polymer waveguide structures can even be fabricated on the tip of an optical fiber \cite{Markiewicz.2019}.
\\
The waveguides can be configured to feature extended planar networks lying on the substrate with three-dimensional features, \textit{e.g.}, for perpendicular coupling.
We harness the optical waveguide both to address and to bidirectionally detect the NV center ensemble.
For addressing, a green laser beam for optical initialization and readout of the spin ensemble is guided through the waveguide to the nanodiamond.
For detection, the fraction of total fluorescence, which is coupled to the waveguide,  is guided to the microscope's focal plane, where input and output ports of the waveguide are imaged simultaneously \cite{Landowski.2017}.
\\
Such integrated NV-center ensembles embedded in polymer waveguides might be used to sense magnetic field gradients and distributions in real time.
Further, the integration of NV centers into a waveguide might enhance the applicability of these color centers as probes in biological tissue, which in general heats, gets photodamaged, and autofluoresces when illuminated with focused green laser light \cite{Alkahtani.2018}.
Since the waveguide, made from a biocompatible material \cite{Hessler.2017}, guides the excitation light and separates it spatially from the biological sample, the phototodamage is strongly reduced in contrast to confocal microscopy, where the excitation light is focused through the sample volume.
\\
We present the integration of strong fluorescent nanodiamonds into a photoresist and the direct laser writing of three-dimensional waveguides with micrometer-scale cross section from this functionalized photoresist.
These devices are characterized with respect to coupling efficiency from nanodiamond to guided waveguide modes, their use as waveguide coupled magnetometer, and influence of waveguide integration on the NV centers' coherence properties.
\section{Experimental System}
\begin{figure*}
	\includegraphics[width=1\textwidth]{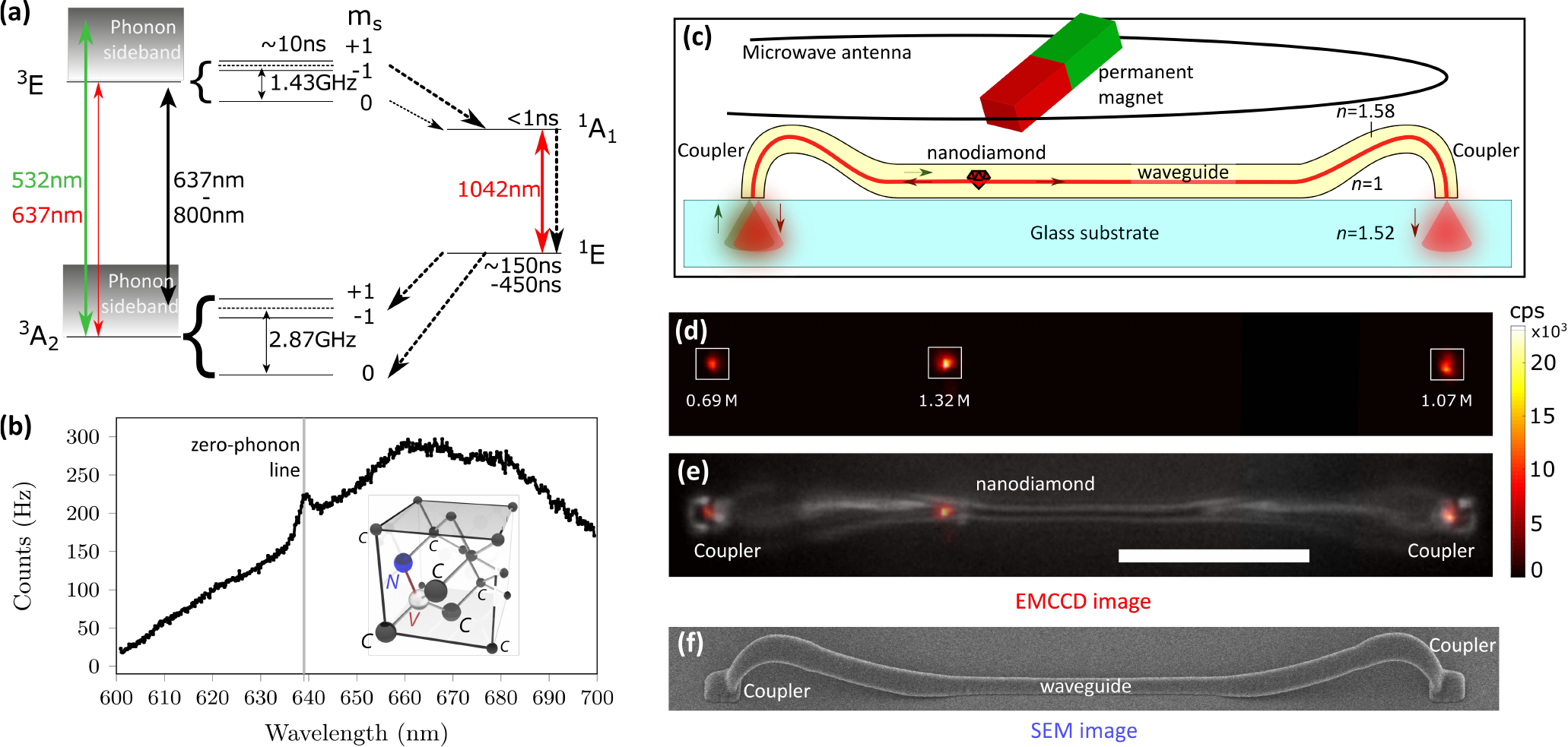}			
	\caption{
		(a) Level structure of the NV center  \cite{Jensen.2014,Acosta.2010c,Rogers.2015}, (b) fluorescence spectrum of the negatively charged NV center in the used nanodiamonds, inset:  lattice structure of the NV center, with the substitutional nitrogen atom in blue, the NV-axis in red and the vacancy in silver.
(c) Sketch of the experimental setup. A green laser  beam is coupled into the waveguide, containing a fluorescent nanodiamond.
		The fraction of the fluorescence guided in the waveguide is detected at the waveguide couplers, the unguided fraction is detected at the nanodiamond's position, while green excitation light is blocked using a long-pass filter in the detection beampath.
		This is shown in the EMCCD image in (d) for a single integrated fluorescent diamond, where the detected fluorescence is shown in false colors and the measured countrate in the white areas is given. In (e), the reflected light microscope image is shown in gray scale, overlaid with the detected red fluorescent light from the nanodiamond. Scale bar: \SI{20}{\micro\meter}.
		For comparison, a tilted SEM image of such a waveguide is shown in (f).
}
	\label{NVC_structure&level}
\end{figure*}
The NV center is a paramagnetic point defect in the diamond lattice, see FIG. \ref{NVC_structure&level} (a), (b). Its structure, shown in the inset of FIG. \ref{NVC_structure&level} (b),  is composed of a substitutional nitrogen atom (N) and an adjacent vacant lattice site (V). 
We use the negatively charged state, which binds an additional electron from nearby N donors \cite{Doherty.2013}.
Its ground state is a spin triplet with $D\sim\SI{2.87}{\giga\hertz}$ zero-field splitting between the $m_{\mathrm{s}}=0$ and degenerate $m_{\mathrm{s}}=\pm1$ spin states, quantized along the NV-axis.
This degeneracy is lifted in presence of an external magnetic field. Ground-state splitting and Zeeman shift, caused by the magnetic field component along the NV-axis, can be described by the Hamiltonian \cite{Doherty.2013}
	\begin{widetext}
	\begin{equation}
	\hat{H}= D \left[\hat{S}^2_z-\frac{S(S+1)}{3}\right]+\mu_B g^{||}_{\mathrm{gs}} \hat{S}_z \cdot B_z + \mu_B g^{\perp}_{\mathrm{\mathrm{gs}}} (\hat{S}_x \cdot B_x + \hat{S}_y \cdot B_y),
	\end{equation}
	\end{widetext}
	where $g^{||}_{\mathrm{gs}}=g^{\perp}_{\mathrm{gs}}=g$ is the $g$-factor \cite{Loubser.1978}.
Due to the system's higher probability to undergo an inter-system crossing and subsequent decay from the excited $\mathrm{^3E}$ $m_\mathrm{s}=\pm1$ states to a long-lived singlet state $\mathrm{^1E}$, the system is optically pumped into the $m_\mathrm{s}=0$ ground state under optical excitation.
This causes a higher fluorescence rate of the $m_\mathrm{s}=0$ state,  harnessed in optical spin-state read out.

\section{Nanodiamond integration into photoresist}
In order to immerse nanodiamonds into the wave\-guides, we follow a probabilistic approach. 
The wave\-guides used are fabricated via direct laser writing in a polymer photoresist, using a commercial system (Nanoscribe Photonic Professional GT \cite{NanoscribeGmbH.2017}).  The nanodiamonds (\SI{120}{\nano\meter} size on average and  more than 1200 NV centers per diamond, Sigma-Aldrich, '798088') were used as received and have been dispersed within the resist EpoClad 50 (micro resist technology).
For technical details on the integration, see Appendix \ref{integration}.
\\
Since the resist film is solid, the nanodiamonds are not mobile and the resist is exposed to laser lithography from the side of the substrate in order to produce waveguides on the substrate, including nano diamonnds.
These waveguides written have been characterized in detail elsewhere \cite{Landowski.2017}. In short, they feature bend radii down to \SI{40}{\micro\meter}, insertion losses (\textit{i.e.} the total loss of the device, including coupling loss on both sides, propagation loss, and bend loss)  on the order of \SI{10}{\decibel}, and show loss coefficients smaller than \SI{0.81}{\decibel\per\milli\meter}. 
To insert and extract light to and from the waveguide, we use three-dimensional out-of-plane couplers with a bend radius of \SI{7}{\micro\meter}.
This small coupler radius is chosen as a compromise between insertion loss, small footprint, and stability.
\\
The number of nanodiamonds integrated into individual waveguides follows a Poissonian distribution \cite{Gutsche.2019}. We select waveguides containing exactly one nanodiamond in the planar waveguide section for the following investigations.
In FIG. \ref{NVC_structure&level} (c-f), a sketch of the waveguide and its working principle are shown, as well as a light microscope image of a waveguide containing one single nanodiamond and a SEM image of a fabricated structure.

\section{Extracting the waveguide coupling efficiency}
\begin{figure}
	\begin{center}
		\includegraphics{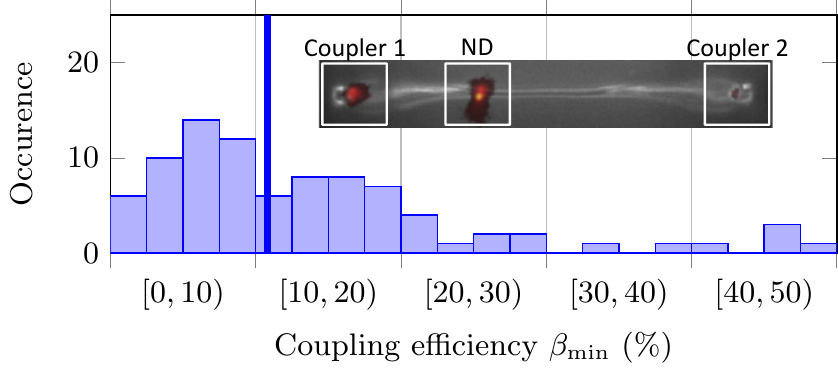}	
	\end{center}
	\caption{Distribution of corrected single-sided coupling efficiencies of the total emission to one propagation direction. 
	The excitation laser is coupled into one of the waveguide ports, the fluorescence counts in the camera image are integrated over a small area surrounding the respective coupler or nanodiamond ('ND').
	The median is marked in blue, negative values were neglected. }
	\label{hist_coupling}
\end{figure}

In order to excite integrated NV centers, a green laser beam ($\lambda=\SI{532}{\nano\meter}$) is focused onto the incoupling port of the waveguide. We use a high numerical aperture objective (Nikon CFI Plan Apochromat Lambda 60XC, $\mathrm{NA}=0.95$), but this can be replaced by other means of coupling to the waveguide, for example an optical fiber.
For fluorescence readout, an EMCCD-camera (Andor iXon 885) or a fiber-coupled single photon counting module (LaserComponents Count 100C-FC) is used.
\\
As can be seen from FIG. \ref{NVC_structure&level} (d), if a nanodiamond is excited by green light through the waveguide it will emit light both into free space and into the waveguide.
An important question regards the coupling efficiency $\beta$ of fluorescence light into the waveguide.
We define the minimum coupling efficiency
\begin{equation}
\beta_{\mathrm{min}}=\frac{\alpha}{\gamma}\cdot\frac{\mathcal{I}_\mathrm{coupler}}{\mathcal{I}_\mathrm{ND}}, \label{eq:efficiency}
\end{equation}
where $\mathcal{I}_\mathrm{coupler}$ ($\mathcal{I}_\mathrm{ND}$) is the fluorescence counts of the nanodiamond guided by the waveguide (emitted into free space) and simultaneously detected by an EMCCD camera; $\alpha=2.85$ corrects for the independently measured coupling loss of the waveguide coupler \cite{Landowski.2017}, and $\gamma=9$ corrects for the limited numerical aperture (NA) of $\mathrm{NA}=0.95$ of the imaging system collecting the free-space fluorescence of the nanodiamond. This correction factor does not apply to the couplers, since the waveguide couplers have $\mathrm{NA}<0.95$.
This yields a lower bound for $\beta$, neglecting the propagation loss of the waveguide and anisotropic emission of the nanodiamond.
Details on the measurement of the coupling efficiency are given in Appendix \ref{coupling}.
\\
We have verified that the fluorescence light originates from the nanodiamond rather than waveguide autofluorescence by modulating the fluorescence rate of the NV center via applying a microwave at the electronic spin resonance (ESR) at \SI{2.865}{\giga\hertz} without an external magnetic field.
In case of smaller ensembles and single NV centers, distinction between the NV center's fluorescence and the waveguide's autofluorescence is feasable taking their different lifetimes into account, see Appendix \ref{waveguide}.
Evaluating the single-side coupling efficiency for both couplers of fifty nanodiamond-containing waveguides separately, we find a distribution of coupling efficiencies (see FIG. \ref{hist_coupling}) with a median of  $\beta_{\mathrm{min}}=10.8\substack{+15.1\\-~5.6}\,\si{\percent}$ into each propagation direction, where the errors denote the standard deviation, corresponding to a total coupling efficiency of $\beta_{\mathrm{min, total}}=21.6\substack{+21.4\\-~7.9}\,\si{\percent}$.
\\
Additionally,  simulations in Comsol Wave Optics \cite{Comsol.} have been performed for different positions of the nanodiamond within the waveguide cross section, see Appendix \ref{simulation}. From the simulations, the maximum coupling efficiency of the integrated emitters to the waveguide mode, \textit{i.e.}~the $\beta$-factor, is  around $\SI{15}{\percent}$ for one propagation direction, which corresponds to $\beta_{\mathrm{max, total}}$ around $\SI{30}{\percent}$ in total.  This value, however, does not include any experimental imperfections, such as finite solid detection angle or coupling and propagation loss of the waveguides. 
Further, it is highly dependent on the nanodiamond's position relative to the mode structure inside the waveguide.
For comparison, from a planar  diamond membrane, only approximately \SI{8}{\percent} of the emitted photons would be collected by a standard microscope objective with NA$=1.3$. 	
To increase this extraction efficiency from bulk diamond, elaborate nanostructuring schemes are necessary, also requiring excellent alignment to the investigated color center.
For example, for a solid immersion lens, up to \SI{29.8}{\percent} coupling efficiency was reported \cite{Hadden.2010} and for parabolic reflector structures up to \SI{41}{\percent} \cite{Wan.2018}.
\\
From the comparison of free-space fluorescence and the fluorescence guided by the waveguide we also find that the ESR fluorescence contrast is only little affected by the waveguide, hence the contribution of waveguide fluorescence is negligible for the NV center concentration used.
\section{Magnetometry}
\begin{figure}
	\begin{center}
		\includegraphics{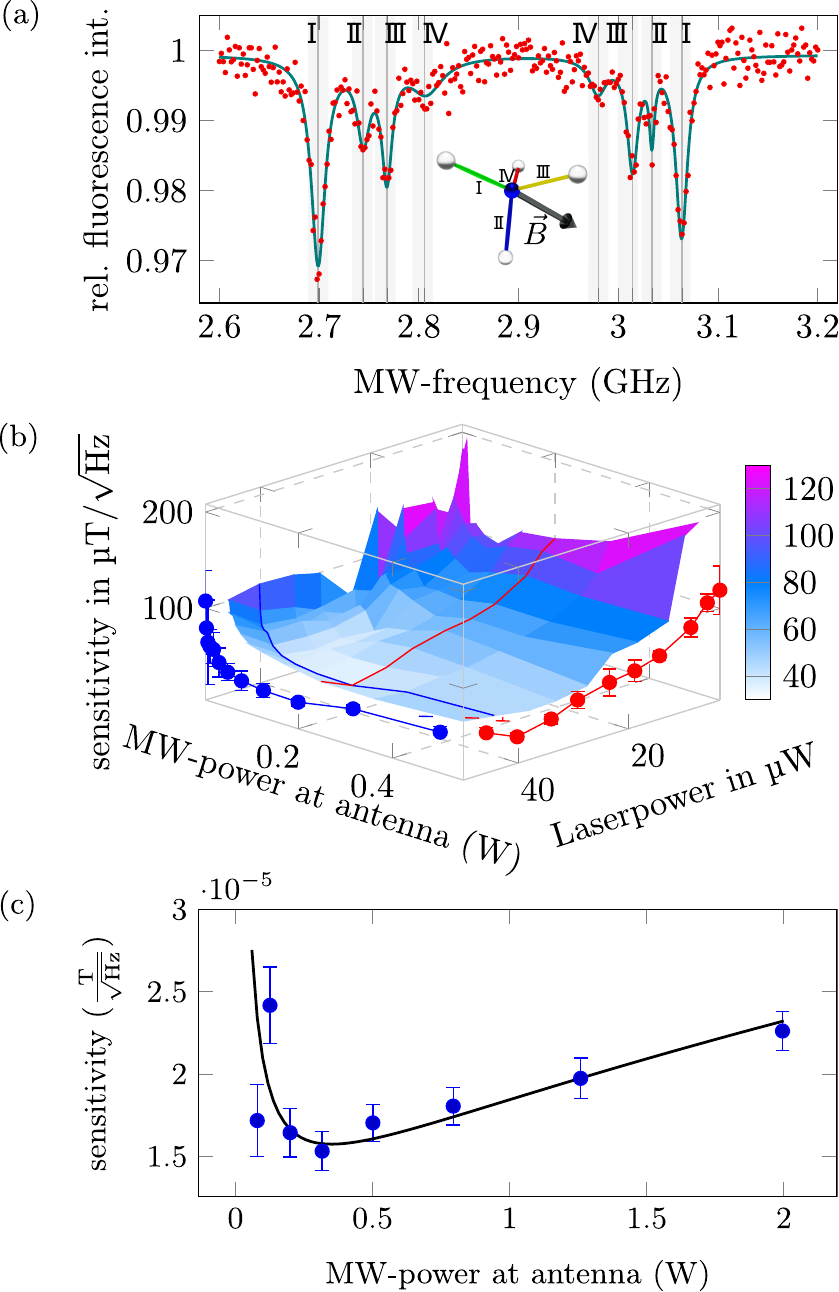}	
	\end{center}
	\caption{(a) ODMR measurement on a nanodiamond detected via the waveguide.
	The inset shows the magnetic field orientation $\vec B$ in gray, relative to the four
		NV center directions, retrieved from the fit data (a sum of eight Lorentzians) shown as solid line.
		The resonances are highlighted in the color of the corresponding NV axis. The magnetic field amplitude is measured to \SI{16.1}{mT} from the ESR positions. 
		(b) Systematical investigation of the dependence of the magnetometer sensitivity on laser and microwave power. Cuts through the minimum for constant laser (red) and microwave power (blue) are shown projected on the faces of the encapsulating box of the plot.
		(c) Calculated cw sensitivity of the NV centers used, depending on the applied MW-power.
The solid curve is a fit to the data according to the formula given in the text with fit parameters $a=0.1721$, $b=\SI{0.03232}{\hertz^2}$, $C=\SI{46.3e-5}{\tesla\per\sqrt{\hertz}}$.
}
	\label{ODMR_spcm}
	\label{sensitivity}
\end{figure}

A possible limitation for application of NV centers as magnetometers is the need for direct optical access to the NV centers under investigation, when probing, for instance, strongly scattering, opaque media, or sample being sensitive to the excitation or fluorescence light.
We show that our waveguides mediate this access to individual integrated nanodiamonds for excitation light and read out, such that these nanodiamonds can be used for vector magnetometry, according to \cite{Kitazawa.2017}.
Another possibility to extract the magnetic field components in a Cartesian basis is to modulate magnetic offset fields in the respective directions with incommensurable frequencies, see \cite{Schloss.2018}.
\\
The position uncertainty of the waveguide-coupled magnetometer is on the order of \SI{1}{\micro\meter} in the waveguide's transversal direction, defined by the waveguide's cross section, and on the order of \SI{0.5}{\micro\meter} along the waveguide axis, determined by the resolution of the microscope used for initial characterization.
In fact, high resolution microscopy is only necessary for initial characterization, all other measurements rely on the use of photon counting devices and the signal extracted from the waveguide couplers, only.
\\
To acquire fast ODMR spectra with high SNR, the microscope is operated in confocal configuration, detecting the fluorescence power with a single photon counting module (SPCM). 
The high resolution of the microscope is not used to detect the nanodiamond's position, but to collect emitted fluorescence photons with high efficiency.
In the current setup, this is only possible for one spot under investigation at a given time.
Such an ODMR spectrum measured via the waveguide is shown in FIG. \ref{ODMR_spcm} (a). 

	In this spectrum, the four pairs of resonances show different amplitudes and linewidths, depending on the polarization of optical and microwave driving fields with respect to each NV-axis orientation \cite{Epstein.2005,Alegre.2007,Herrmann.2016}. The resonances' amplitudes also differ between detection of the non-guided and guided waveguide modes.
	A comparison of ODMR-spectra for guided and non-guided modes is shown and further discussed in  Appendix \ref{waveguide}. Observation of such ODMR spectrum at the waveguide coupler confirms the presence of an individual singlecrystalline nanodiamond inside the waveguide, because a second nanodiamond would contribute additional resonances due to deviations of its crystal orientation.
Further, the magnetic field orientation with respect to the diamond lattice is determined from this spectrum, see the inset in FIG. \ref{ODMR_spcm} (a), and the magnetic field amplitude is determined to be \SI{16.1}{mT}. 
\\
The figure of merit for a magnetometer is its sensitivity.
For cw-ODMR measurements, the shot noise limited sensitivity is given by \cite{Dreau.2011}
\begin{equation}
 \eta_B=\mathcal{P}\frac{h}{g\mu_{\mathrm B}}\frac{\Delta\nu}{\mathcal{C}\sqrt{\mathcal R}}
 \label{eq:sensitivity}
\end{equation} 
where $\mathcal{P}\approx 0.77$ for a Lorentzian resonance shape, $\Delta \nu$ is the FWHM of the dip, $\mathcal C$ is the dip depth, and $\mathcal R$ is the detected count rate.
The dependence of the sensitivity on laser and microwave power for one waveguide coupled nanodiamond is systematically investigated in FIG. \ref{ODMR_spcm} (b). 
Over a broad range of laser and microwave powers, the sensitivity varies only slowly and shows a maximum for low microwave and moderate laser powers. It decreases dramatically for very low microwave and laser powers.
Maximum sensitivity is found for a different nanodiamond, as shown in FIG \ref{ODMR_spcm} (c).
There, we determine the sensitivity of the waveguide-coupled magnetometer for a fixed laser power by measuring the width and depth of the dip at lowest frequency while varying the microwave power, see FIG. \ref{sensitivity}.
We fit the sensitivity with a function
of the form $\eta_B=C\cdot\sqrt{a+\frac{P}{b}}/(\frac{P}{P+a \cdot b})$,  which is a simplified form of the theoretical description in \cite{Dreau.2011} for constant excitation laser power.
Typical values are $\mathcal C \approx$ \SIrange{0.014}{0.027}{}, $\mathcal R \approx 800000$, $\Delta \nu\approx$\SIrange{7}{20}{\mega\hertz}.
The maximum sensitivity  is \SI[separate-uncertainty = true]{1.53(12)e-5}{\tesla\per\sqrt{\hertz}} at a microwave power of \SI{0.316}{\watt} applied to the wire loop antenna. 
The microwave power can not easily be compared for both devices shown, since the coupling strength of microwave to nanodiamond is highly dependent on the position of the antenna.
For higher microwave powers the sensitivity decreases, since the width of the resonance is dominated by power broadening of the microwave, while for lower microwave powers the decrease of the sensitivity is possibly dominated by laser power broadening of the resonance due to optical polarization of the spin state, shot noise, or spin projection noise.\\
To circumvent power broadening, a pulsed measurement scheme, temporarily separating microwave and laser pulses could be used \cite{Dreau.2011}.
For the nanodiamonds in our case, however, the sensitivity is inherently limited by the diamonds' quality.
In fact, the improved sensitivity of sensors made of ultra-pure bulk samples is due to a  reduced abundance of $^{13}$C-atoms, lower amount of nitrogen (P1-center or $\mathrm{N_s^0}$) and other impurities, acting as spin bath and reducing NV-center's coherence times \cite{Balasubramanian.2009,Barry.2019}. 
\section{Coherent Spin Dynamics}
\begin{figure}
	\begin{center}
				\includegraphics{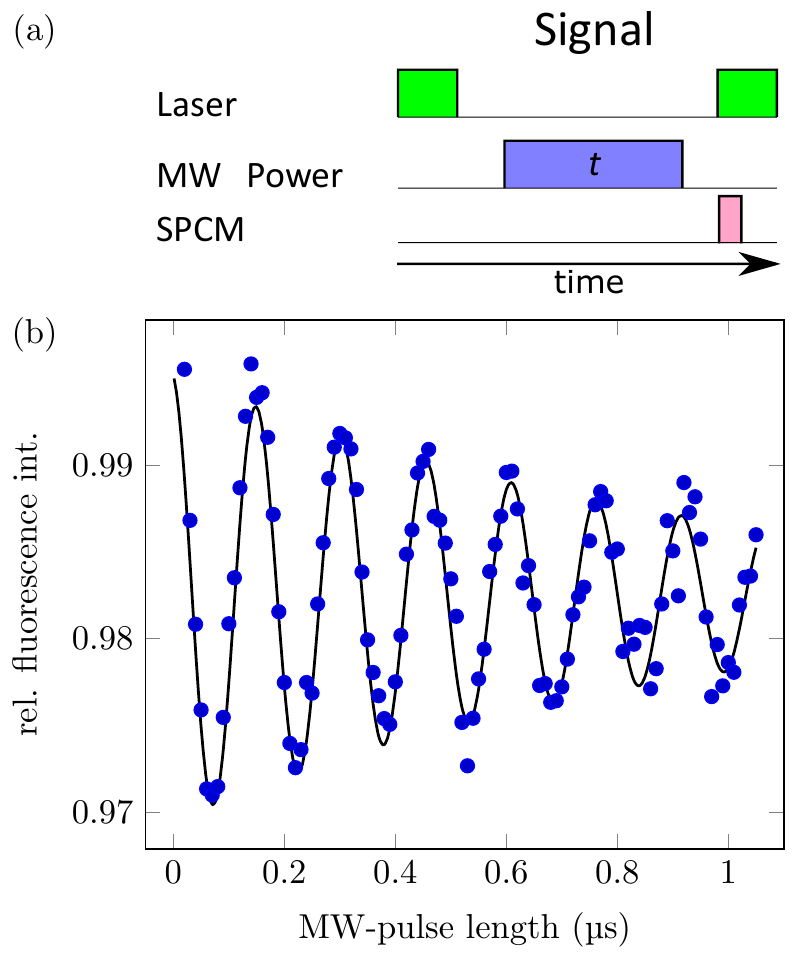}	
	\end{center}
	\caption{
		(a) Experimental sequence to measure Rabi oscillations,
		(b) Rabi oscillations of NV centers in a single nanodiamond, detected via the waveguide. \SI{2}{\watt} MW power at the antenna. The solid line is a fit to a damped sine, yielding a damping time of $T\approx\SI{0.9}{\micro\second}$.
	}
	\label{rabi}
\end{figure}

A remarkable property of the NV center is its relatively long room-temperature spin-coherence time, which can reach $T_2=\SI{3}{\milli\second}$ \cite{BarGill.2013}, depending on the diamond's quality.
It is known that for nanodiamonds the surface can have profound impact on the embedded NV-centers' coherence times as in the case of the present nanodiamonds, used without surface cleaning steps. 
Therefore we verify that integrating the nanodiamonds into the photoresist does not alter the NV centers' coherence time.
We show the remote detection of Rabi oscillations of an ensemble of NV centers through an optical waveguide, see FIG. \ref{rabi}, where the coherence time is compatible to the coherence time of uncoupled nanodiamonds, limited by the impurities in the volume and on the surface of the nanocrystals.
Without waveguide, we measure typical coherence times of $T_2\approx\SI{1.5}{\micro\second}$ and $T_2^*\approx\SI{78}{\nano\second}$, for a waveguide integrated nanodiamond we measure coherence times of $T_2\approx\SI{1.4}{\micro\second}$ and $T_2^*\approx\SI{74}{\nano\second}$.
The coherence times depend strongly on the nanodiamonds quality and hence vary from nanodiamond to nanodiamond.
For details on the measurement of Rabi oscillations, Hahn-Echo, and Ramsey measurements, please see  Appendix \ref{rabi_measurement}.
\\
The Rabi oscillations detected show a coherence time of \SI{0.9}{\micro\second}, which is comparable to coherence times measured in bare nanodiamonds. 
We expect that reducing the number of surface impurities and lowering the N-concentration of the nanodiamonds embedded into the waveguides will enhance coherence times \cite{Tisler.2009,Knowles.2014}.
\\
\section{Conclusion}
We have shown the integration of single crystalline nanodiamonds containing an ensemble of NV centers into direct-laser-written waveguides, which enables the creation of extended three-dimensional waveguide networks on a chip. 
These NV centers can be used as integrated sensors for magnetic fields, which are initialized and read out via the waveguides they are embedded in. 
The ODMR signal of the NV ensemble is not affected by fluorescence of the waveguide material. However, for small ensembles or even single NV centers the fluorescence contributions of different origin might be separated by harnessing the strongly different lifetimes of  \SI{33.6}{\nano\second} (\SI{5.28}{\nano\second})  for NV centers (photoresist), see Appendix \ref{waveguide}.
Further, we have shown  the detection of Rabi oscillations via the waveguide housing the nanodiamond, allowing for coherent control over an ensemble of remote quantum emitters. 
The $T_2^*$ time of the waveguide embedded NV ensembles is relatively short ($T_2^*\approx\SI{80}{\nano\second}$),  probably limited by the large number of impurities in and on the surface of our nanodiamonds. While this limits their use for magnetometry using a Ramsey scheme, further enhancement can be achieved by employing nanodiamonds with improved coherence properties.
	Thereby, our waveguides pave the way to optical networks, hosting three-dimensional arrays of spin sensors.
In the future it will be interesting to explore the prospects of  detecting magnetic fields as well as gradients in a small volume by direct-laser-written three-dimensional photonic structures with multiple nanodiamonds embedded.
Another option is to directly connect the waveguide to an optical fiber, where no confocal microscope is needed any more.
\\
These sensors are also useful for opaque, scattering, or sensitive  biological samples under investigation, since the laser light used to excite NV centers, as well as their fluorescent light, is guided within the waveguide, reducing the amount of optical power deposited at the sample by orders of magnitude and simultaneously keeping the whole sample optically accessible. At the same time, the processed photoresist is stable against watery solutions and might thus be compatible with sensing applications in life science.
\begin{acknowledgments} 
	We thank Henning Fouckhardt and his group Integrierte Optoelektronik und Mikrooptik (Technische Universit\"at Kaiserslautern) for access to their wet chemistry lab, and we also acknowledge the technical support by the Nano Structuring Center (Technische Universit\"at Kaiserslautern).
	Furthermore, we thank Stefan Dix for help with the experimental setup.
\end{acknowledgments}
\bibliography{bibliography}
\appendix
\section{Nanodiamond integration into photoresist}
\label{integration}
	Since the nanodiamonds are received suspended in deionized water (concentration: \SI{1}{\milli\gram\per\milli\litre}),
	the deionized water is exchanged by gammabutyrolactone (GBL), the solvent of EpoClad. 
	This is achieved by, first, centrifugating the suspension for \SI{15}{\minute} at \SI{6000}{rpm}, and second, exchanging the water by the same amount of GBL.
	To suspend the new mixture, it is vortexed and exposed to an ultra sonic bath to disperse agglomerates into single crystalline nanodiamonds.
	This procedure is repeated two times to ensure a low residual water concentration.
	For integration of nanodiamonds into the photoresist,
	equal mass ratios of the suspension and EpoClad 50 are mixed for \SI{30}{\minute} at \SI{200}{rpm}, using a magnetic stir bar. 
	The photoresist with immersed nanodiamonds is processed for direct laser writing as described in Ref.~\cite{Landowski.2017}.
\section{Waveguide coupling efficiency}
\label{waveguide}
\begin{figure}
	\begin{center}
		\includegraphics{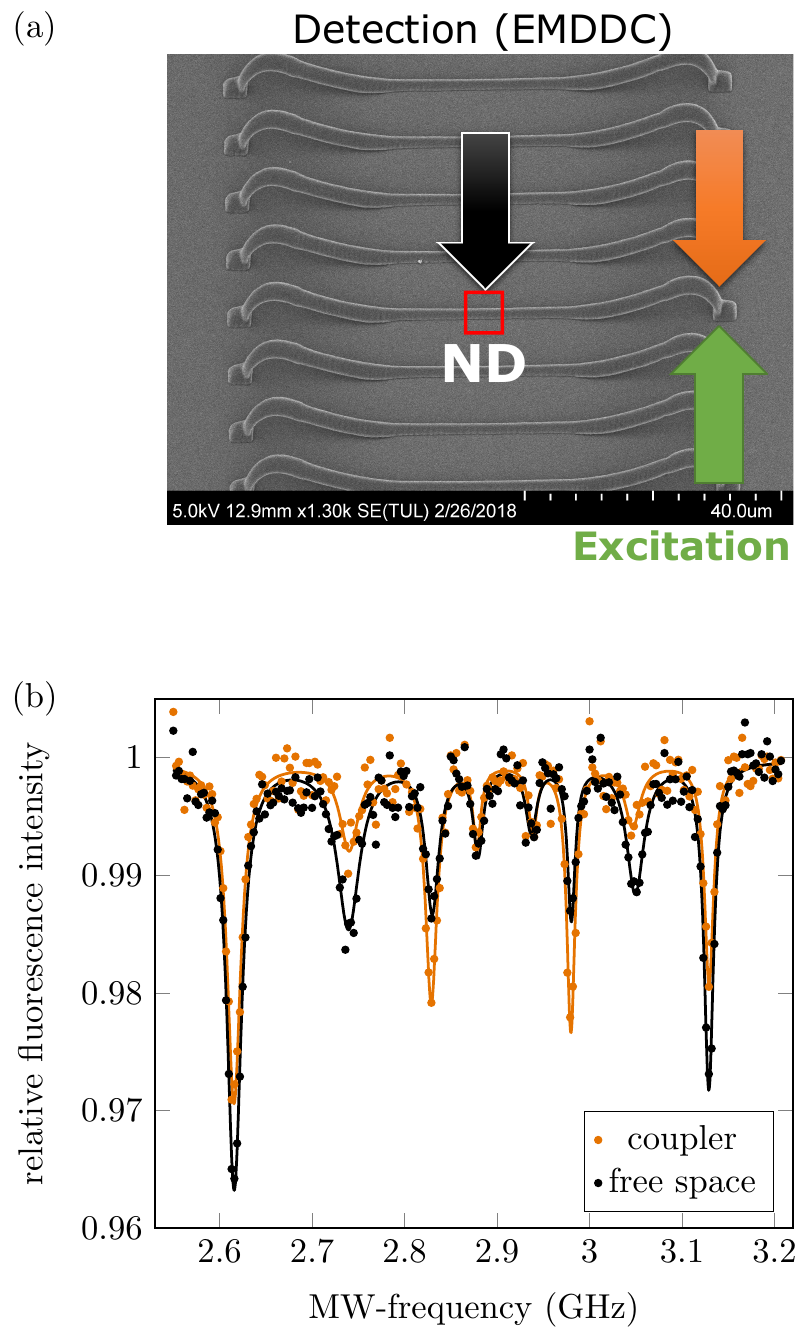}	
	\end{center}
	\caption{(a) SEM image of series of direct-laser-written waveguides with nanodiamond indicated by square. (b) Comparison of ODMR signals excited via the waveguide and detected via waveguide (orange), free space (black).}
	\label{ODMR_cam}
\end{figure}

\begin{figure}
	\begin{center}
		\includegraphics{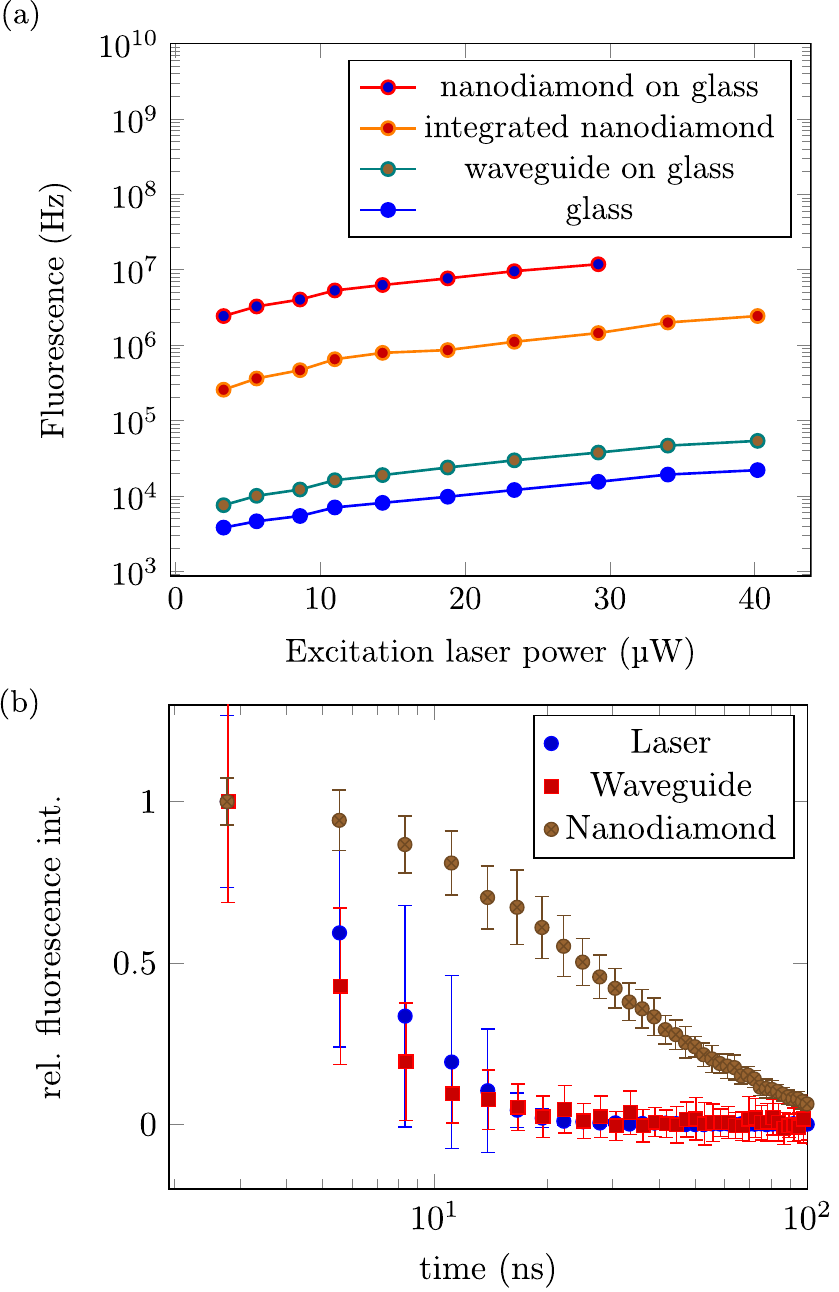}		
	\end{center}
	\caption{
	(a) Detected fluorescence count rates for bare nanodiamond on glass, an individual integrated nanodiamond, waveguide on glass, and glass coverslip, (b) fluorescence lifetimes of the photoresist measured by exciting a waveguide in comparison to lifetime of NV centers in nanodiamond and the laser switching characteristics. The fluorescence signal of the resist is much noisier due to the low countrate (ratio of initial countrates is approximately $10^2$), the fluorescence intensity of each data set is independently normalized to 1 in the beginning. The errorbars are statistical ($1\sigma$) errors.
	}
	\label{FLIM}
\end{figure}

\begin{figure}
		\begin{center}
			\includegraphics[width=0.5\textwidth]{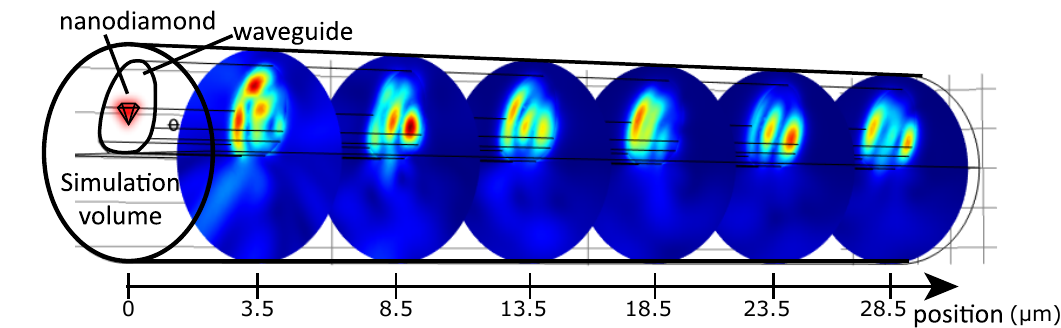}
		\end{center}
		\caption{Simulated propagating light field in the waveguide, emitted by a nanodiamond positioned at the center of the fundamental mode. From the ratio of guided energy to total irradiated energy the coupling factor can be deduced.
		}
		\label{simulation_image}
	\end{figure}

In order to compare the ODMR signal collected directly from the nanodiamond at the one hand and via the waveguide at the other hand, an ODMR measurement is taken. Using the microscope's EMCCD camera both the waveguide port and the nanodiamond's free-space fluorescence are imaged simultaneously. 
The resulting spectra are shown in FIG. \ref{ODMR_cam}, showing the spin-resonances, as expected, for the same frequencies. 
However, the contrast of specific dips differs for guided and non-guided modes detected via the coupler and at the nanodiamond's position, respectively. We attribute this to different overlap of the two transition dipole orbitals \cite{Epstein.2005,Chapman.2011,Kaiser.18.06.2009} of each NV-orientation to the guided modes.
\\
In the case presented here, on the order of $10^3$ NV centers contribute to the signal. Neglecting the influence of the polymer waveguide might thus not be valid for weaker signals of few or single NV centers.
This is shown in FIG \ref{FLIM} (a) for a comparison of detected count rates of the components of the material system used in dependence of excitation laser power, which shows that a different substrate might improve  SNR for a lower concentration of NV centers.
 In this case, the contribution of NV fluorescence versus waveguide fluorescence may still be separated by exploiting markedly different time scales of fluorescence lifetimes in a pulsed measurement, as shown in FIG.~\ref{FLIM} (b). 
\label{coupling}
	\begin{figure}
	\begin{center}
		\includegraphics{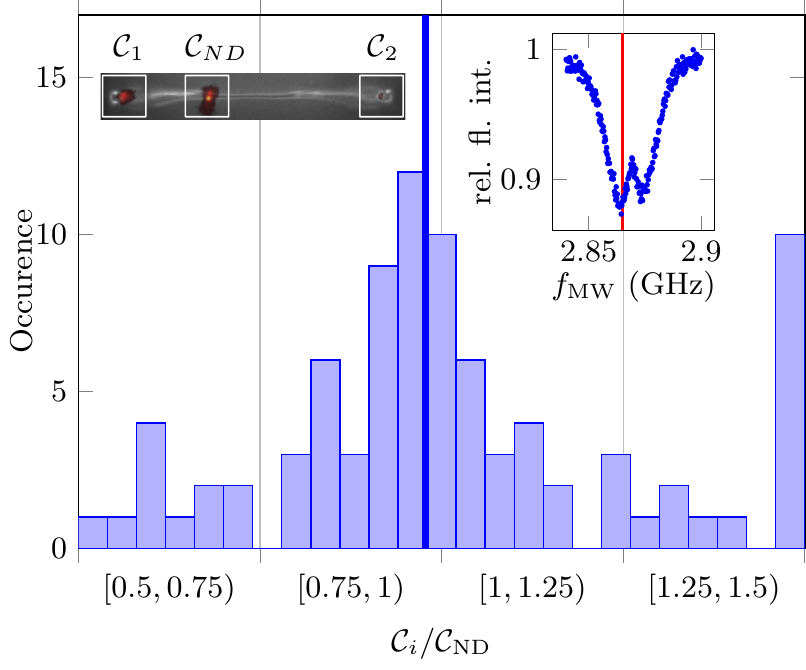}		
	\end{center}
	\caption{
		Distribution of ratio of  ESR contrasts $\mathcal C_i$ at \SI{2.865}{\giga\hertz}, detected at the couplers $i=1,2$  relative to the contrast $\mathcal C_{\mathrm{ND}}$, detected at the nanodiamond. The median is marked in blue. The contrast $\mathcal C_i$ is defined as $\mathcal C_i=1-\frac{\mathcal I_i'}{\mathcal I_i}$, where $\mathcal I_i'$ ($\mathcal I_i$) is the detected count rate at position $i$ with (without) microwave applied.
		(insets) Free space ODMR spectrum in absence of an external magnetic field, the microwave frequency used is marked in red; microscope image of a waveguide under investigation with marked positions for evaluation.
	}
	\label{hist_dip}
\end{figure}
\\
	We prepare and measure 50 nanodiamonds, each embedded in another waveguide structure. 
	In order to verify that the fluorescence light originates from the nanodiamond rather than waveguide autofluorescence, we modulate the fluorescence rate of the NV center by applying a microwave at the electronic spin resonance (ESR)  at  \SI{2.865}{\giga\hertz} (vertical line in the inset of FIG.~\ref{hist_dip}) without an external magnetic field. The distribution of ratios between both contrasts is shown in FIG.~\ref{hist_dip} and it is well centered around unity  (median of 0.978).
	We conclude that for our concentration of NV centers in the nanodiamond the fluorescence contrast is only little affected by the waveguide, which shows that the contribution of waveguide fluorescence is negligible for the  NV center concentration used.
\\
	Experimentally, we determine the quantities in Eq.~(\ref{eq:efficiency}) to extract the minimum coupling efficiency $\beta_{\mathrm{min}}$.
	Evaluating camera images, we can simultaneously determine the fluorescence of the nanodiamond emitted into free space $\mathcal{I}_\mathrm{coupler}$ and the guided fluorescence at the waveguide output couplers $\mathcal{I}_\mathrm{ND}$. 
	\\
	Comparing the ODMR-signal counts at the waveguide couplers with those detected from free space, corrected by $\gamma$, we get a median bare ratio of the counts of
	$\mathcal{I}_\mathrm{coupler} / (\gamma \cdot \mathcal{I}_\mathrm{ND})=3.8\substack{+5.3\\-~2.0}\,\si{\percent}$  for the coupling efficiency of total emission to the waveguide modes propagating into one direction, with error given by standard deviation, see FIG. \ref{hist_coupling}.
	If we take the correction factor $\alpha$  into account, we get a corrected median coupling efficiency of $\beta_{\mathrm{min}}=10.8\substack{+15.1\\-~5.6}\,\si{\percent}$.
	This spread is caused by a strong dependence of the coupling efficiency on the position of the nanodiamond in the waveguide.
	\\
	Additionally, we performed simulations in Comsol Wave Optics to estimate the maximum coupling efficiency and coupling to the waveguide modes. In FIG. \ref{simulation_image}, the intensity distribution of the emitted light from a fluorescent nanodiamond is shown at multiple propagation distances for one of those simulations.
	\label{simulation}
\section{Measurement of Rabi oscillations and coherence times}
\label{rabi_measurement}
\begin{figure}
\begin{center}
	\includegraphics{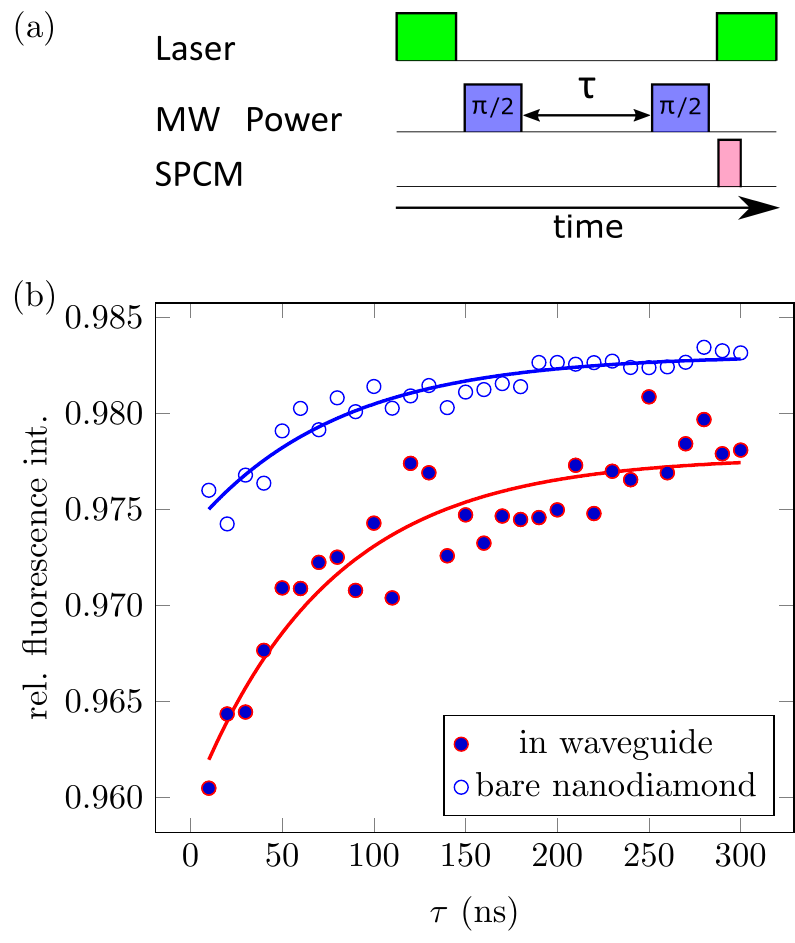}	
\end{center}
\caption{(a) Experimental sequence for Ramsey measurements, (b)
	Comparison of Ramsey measurements for a nanodiamond inside a waveguide and  bare nanodiamond on the glass coverslip. The coherence time $T_2^*$ was extracted from an exponential decay fit to $T_2^*=\SI[separate-uncertainty=true]{74(15)}{\nano\second}$ ($T_2^*=\SI[separate-uncertainty=true]{78(13)}{\nano\second}$) for the nanodiamond in the waveguide (bare nanodiamond).
The length of the $\pi/{2}$ pulses is \SI{30}{\nano\second}.
}
\label{ramsey}
\end{figure}

\begin{figure}
	\begin{center}
		\includegraphics{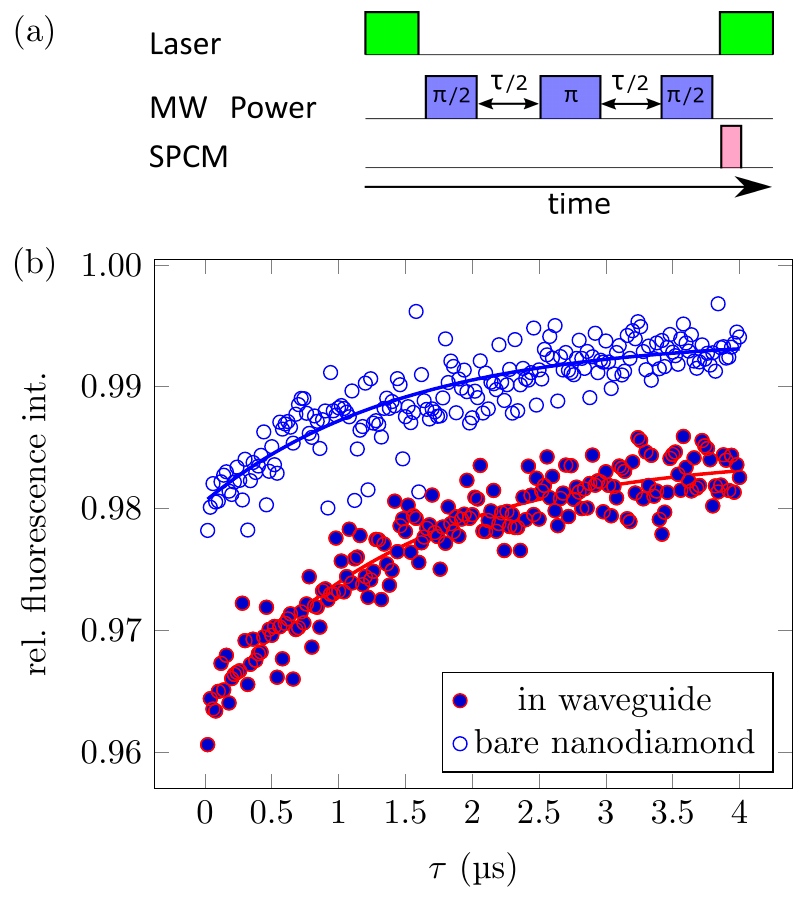}
	\end{center}
	\caption{(a) Experimental sequence for Hahn-Echo measurements, (b)
		Hahn-Echo signal for a nanodiamond inside a waveguide and bare nanodiamond on the glass coverslip. The coherence time $T_2$ was extracted from an exponential decay fit to 
		$T_2=\SI[separate-uncertainty=true]{1.4(1)}{\micro\second}$ ($T_2=\SI[separate-uncertainty=true]{1.5(2)}{\micro\second}$) for the nanodiamond in the waveguide (bare nanodiamond). 
		In the case of the waveguide embedded nanodiamond, ${\pi}/{2}$  ($\pi$) pulses of \SI{30}{\nano\second} (\SI{60}{\nano\second}) duration were used.
}
	\label{hahn}
\end{figure}

	In order to address only one single spin transition of one NV orientation of the ensemble, an external magnetic field is applied and an ODMR spectrum recorded.
	A well separated ESR with dip depth of around \SI{3}{\percent}  is chosen for the microwave frequency.
	The Rabi measurement sequence consists of three pulses.
	First, the spin ensemble is initialized in the $m_{\mathrm s}=\ket{0}$ 
	state via a green laser pulse of \SI{50}{\micro\second} duration, coupled to the waveguide.
	Second, the microwave pulse of varied length is applied transferring the addressed ensemble spins into the state $m_{\mathrm s}=\ket{-1}$ or $\ket{+1}$.
	Subsequently, a green laser pulse for state readout is coupled to the waveguide, similar to initialization.
	The fluorescence light emitted during the first \SI{1.5}{\micro\second} of the readout pulse of \SI{3}{\micro\second} duration is guided to an SPCM  via the waveguide and the confocal microscope. The detected number of photons is normalized to the number of photons detected without the microwave field applied, and it shows oscillation between the two spin states.
	The excitation pulses are chosen relatively long compared to single emitters to protect the SPCM from too intense illumination due to the large NV ensembles investigated.
	In order to remove effects from drifts, or fluctuations of, \textit{e.g.}, surface charges and charge states of crystal defects, this sequence is followed by the same sequence without microwave pulse, which is used as reference for normalization.
\\
From the Rabi oscillations, the duration for $\pi/{2}$ and $\pi$ pulses are determined and used to determine $T_2^*$ and $T_2$ times from Hahn-Echo and Ramsey measurements, see FIG. \ref{ramsey}, \ref{hahn}, respectively.
These measurements are quite similar to the measurement of Rabi oscillations, besides the fixed length of the applied microwave pulses and varying precession time between the microwave pulses.
\end{document}